\documentstyle[12pt,titlepage]{article}
 \input epsfig.sty

.5 scaled \magstep4
\font\medio=cmr9.5 scaled \magstep2
\outer\def\beginsection#1\par{\medbreak\bigskip
      \message{#1}\leftline{\bf#1}\nobreak\medskip
\vskip-\parskip
      \noindent}

\def\laq{\raise 0.4ex\hbox{$<$}\kern -0.8em\lower 0.62
ex\hbox{$\sim$}}
\def\gaq{\raise 0.4ex\hbox{$>$}\kern -0.7em\lower 0.62
ex\hbox{$\sim$}}

\begin{document}
\bibliographystyle {unsrt}

\titlepage

\begin{flushright}
CERN-PH-TH/2007-259
\end{flushright}

\vspace{15mm}
\begin{center}
{\Large A magnetized completion of the $\Lambda$CDM paradigm}\\
\vspace{15mm}
 
Massimo Giovannini$^{a,c}$ and Kerstin E. Kunze$^{b,c}$\\

\vskip2.cm

{\sl $^a$Centro ``Enrico Fermi", Compendio del Viminale, Via 
Panisperna 89/A, 00184 Rome, Italy}
\vskip 0.2cm 
{\sl $^b$ Departamento de F\'\i sica Fundamental, \\
 Universidad de Salamanca, Plaza de la Merced s/n, E-37008 Salamanca, Spain}
\vskip 0.2cm 
{\sl $^c$  Department of Physics, Theory Division, CERN, 1211 Geneva 23, Switzerland}
\vspace{6mm}

\end{center}

\vskip 2cm
\centerline{\medio  Abstract}
The standard $\Lambda$CDM paradigm is complemented with a magnetized 
contribution whose effects on the anisotropies of the Cosmic Microwave Background (CMB) 
are assessed by means of a dedicated numerical approach. The accuracy on the
temperature and polarization correlations stems  from the inclusion of the 
large-scale magnetic fields both at the level of the initial conditions and at  the level 
of the Einstein-Boltzmann hierarchy which is consistently embedded in a generalized magnetohydrodynamical 
framework. Examples of the calculations of the temperature 
and polarization angular power spectra are illustrated and discussed. The reported results and the 
described numerical tools 
set the ground for a consistent inclusion of a magnetized contribution in current strategies 
of cosmological parameter estimation.
\noindent

\vspace{5mm}
\vfill 
\newpage
Current analyses of cosmological data sets 
(see, for instance, \cite{data}) are customarily performed using a variety of theoretical models which represent diverse completions (i.e. delicate improvements) of the 
pivotal $\Lambda$ cold dark matter paradigm ($\Lambda$CDM in what follows).
As an example, if we ought to know how large could be the contribution 
of a stochastic background of gravitational waves to the 
anisotropies of the Cosmic Microwave Background (CMB in what follows),
the possible presence of  tensor modes should be added, as a
supplementary feature, to the basic list of cosmological parameters 
of the standard $\Lambda$CDM paradigm whose updated version
 (improved by a tensor contribution) will then be compared with the experimental data. From the latter comparison
 likely values of the ratio between the tensor and scalar power spectra can be inferred. Other possible completions of the $\Lambda$CDM lore might include, for instance, massive neutrinos, a minimal duration of the inflationary phase, an effective barotropic index for the dark energy component and many others. Yet another class of delicate improvements 
of the $\Lambda$CDM paradigm contemplates the inclusion 
of one (or more) non-adiabatic modes which can be either 
correlated or anticorrelated with the standard adiabatic component 
(see, for instance, \cite{h1}).  For specific choices of the non-adiabatic 
amplitude and spectral index the fit to the experimental data may even improve (see, for instance, last reference in \cite{h1}).

In this paper we are going to describe another completion 
of the $\Lambda$CDM paradigm. The possibility we are going to present 
will be dubbed as magnetized $\Lambda$CDM scenario (m$\Lambda$CDM in what 
follows). 
Large-scale magnetic fields arise over different scales ranging from galaxies to clusters \cite{mag1}. 
Superclusters have also been  claimed 
to have magnetic fields \cite{mag1} at the $\mu$G level even if firmer 
evidence is still lacking.
Hopefully  
some of the findings of the Auger  project \cite{auger1} could be used for a magnetic ``tomography" of the local Universe, say within a cocoon of 60 Mpc.

If the present magnetized structures 
emanate from pre-existing cosmological relics they must be present prior to matter-radiation equality, affecting, 
in this way, the physics of photon decoupling and, ultimately, the formation of the CMB anisotropies. 
In the m$\Lambda$CDM scenario 
the CMB anisotropies are computed in the presence of a stochastic 
magnetic field \footnote{A stochastic magnetic field does not 
break, by definition, the overall spatial isotropy of the background 
geometry and allows, as a consequence, for angular 
power spectra just expressed in terms of their dependence 
upon the multipole $\ell$ without any further preferred direction. 
For the opposite situation see, for instance, \cite{dir}.} which will affect both the initial conditions
and the dynamical evolution of the Einstein-Boltzmann hierarchy:
\begin{eqnarray}
&&\langle B_{i}(\vec{k}) B_{j}(\vec{p}) \rangle = \frac{2\pi^2}{k^3} \delta^{(3)}(\vec{k} + \vec{p}) P_{\mathrm{B}}(k)P_{ij}(k),
\nonumber\\
&&P_{ij}(k) = \delta_{ij} 
- \frac{k_{i} k_{j}}{k^2}, \qquad P_{\mathrm{B}}(k) = A_{\mathrm{B}} \biggl(\frac{k}{k_{\mathrm{L}}}\biggr)^{n_{\mathrm{B}}-1}.
\label{E1}
\end{eqnarray}
The minimal m$\Lambda$CDM scenario
contains, on top of the (six) parameters of the $\Lambda$CDM paradigm,  
two new parameters, namely the magnetic spectral index $n_{\mathrm{B}}$ 
and the amplitude of the magnetic power spectrum 
$P_{\mathrm{B}}(k)$ evaluated at the magnetic pivot scale 
$k_{\mathrm{L}}$. Non minimal extensions of the m$\Lambda$CDM scenario include:
the correlation (or the anticorrelation) 
between the adiabatic mode and the magnetic field, the simultaneous presence 
of a magnetized adiabatic mode together with one (or more) magnetized isocurvature 
modes. For sake of simplicity we will stick here to the minimal situation reporting elsewhere on the complementary cases.
There have been lately theoretical and semi-analytical 
works along this direction \cite{mg} but the moment has now come 
to tailor, for the first time, a consistent numerical approach to treat 
the effects of a fully inhomogeneous magnetic field on the scalar modes of the geometry. 

Since, after neutrino decoupling, the concentration of electrons and protons 
is roughly $10$ orders of magnitude smaller than the concentration of the photons, the Debye length-scale will be 
approximately $20$ orders of magnitude smaller than the Hubble radius (for instance at matter-radiation equality). The conductivity of the globally 
neutral plasma, i.e. $\sigma_{\mathrm{c}}$, will be dominated by Coulomb scattering so that $\alpha_{\mathrm{em}} \sigma_{\mathrm{c}}\simeq 4\pi(T/m_{\mathrm{e}})^{1/2} T (\ln{\Lambda_{\mathrm{C}}})^{-1}$ where 
$\Lambda_{\mathrm{C}}$ is the argument of the Coulomb logarithm.
 Ohmic electric fields are then suppressed with respect to the total Ohmic current, as it is the case in a good terrestrial conductor:
 \begin{equation}
 \vec{E} + \vec{v}_{\mathrm{b}} \times \vec{B}= \frac{\vec{J}}{\sigma} \simeq \frac{\vec{\nabla}\times \vec{B}}{4\pi\sigma},\qquad \sigma = a \sigma_{\mathrm{c}},
 \label{E2}
 \end{equation}
where $\vec{E} = a^2 \vec{{\mathcal E}}$ and $\vec{B}= a^2 \vec{{\mathcal B}}$ are the electric and magnetic fields rescaled through the second 
power of the scale factor $a(\tau)$ of a (conformally flat) Friedmann-Robertson-Walker (FRW) geometry. The large-scale description of our problem is, physically, the curved-space version of MHD where the electric field, the magnetic field and the Ohmic current are all solenoidal quantities. The effects of the magnetic field on the scalar modes of the geometry are far from trivial: the evolution of the bulk velocity of the plasma is given by\footnote{A prime denotes a derivation with respect to the conformal time coordinate $\tau$ and ${\mathcal H} = (\ln{a})'$.
The notation $\theta_{X} = \vec{\nabla}\cdot \vec{v}_{X}$ 
denotes the three-divergence of the peculiar velocity of the species $X$. Throughout the paper $\epsilon' = x_{\mathrm{e}} n_{\mathrm{e}} \sigma_{\mathrm{Th}} n_{\mathrm{e}} a/a_{0}$ denotes the optical depth. Finally, with similar notations, $\delta_{X} = \delta_{\mathrm{s}} \rho_{X}/\rho_{X}$ is the density contrast of the species $X$.}
\begin{equation}
 \theta_{\mathrm{b}}' + {\mathcal H} \theta_{\mathrm{b}} = \frac{4}{3} \frac{\rho_{\gamma}}{\rho_{\mathrm{b}}} \epsilon' 
(\theta_{\gamma} - \theta_{\mathrm{b}}) + \frac{ \vec{\nabla} \cdot [ \vec{J} \times \vec{B}] }{a^4 \rho_{\mathrm{b}}}, \qquad \theta_{\mathrm{b}} = \vec{\nabla}\cdot\vec{v}_{\mathrm{b}}.
\label{E3}
\end{equation}
Given the hierarchy between the electron and proton masses the bulk velocity of the plasma $\theta_{\mathrm{b}}$
will be essentially the proton velocity. Because of the 
Coulomb-dominated conductivity, the magnetic fields will be present and not diffused 
for typical length-scales larger than the magnetic diffusivity 
length $L_{\sigma}$ whose ratio to the Hubble radius is so small 
(i.e. $H L_{\sigma} \simeq 3.9\times 10^{-17} (T/\mathrm{eV})^{1/4}$)
that the $k_{\sigma} \simeq L_{\sigma}^{-1}$ will provide effectively an ultra-violet 
cut-off to the magnetic power spectrum of Eq. (\ref{E1}). 
In the presence of diffusion 
damping (i.e. shear viscosity) the Silk wave-number will be the dominant dissipative scale for the baryon-photon fluid. Under these conditions the magnetic flux
(and helicity) will be effectively conserved to a very good approximation.

Large values of the conductivity break explicitly Lorentz invariance. The plasma frame, where the electric fields are suppressed in comparison with the magnetic fields, arises naturally. The scalar fluctuations of the geometry, still relativistic, 
will be treated, numerically,  in the synchronous 
gauge ($S$-gauge in what follows) where the only non-vanishing entries of the perturbed metric are, in Fourier space:
\begin{equation}
\delta_{\rm s} g_{i j}(k,\tau) = 
a^2(\tau)\biggl[ \hat{k}_{i} \hat{k}_{j} h(k,\tau) + 6 \xi(k,\tau)\biggl(  
\hat{k}_{i} \hat{k}_{j} - \frac{1}{3} \delta_{ij}\biggr)\biggr],
\label{pert1}
\end{equation}
where the symbol $\delta_{\mathrm{s}}$ emphasizes that we are dealing here with {\em scalar} (as opposed to vector or tensor \cite{vt}) fluctuations of the geometry. The m$\Lambda$CDM code is an extension of the CMBFAST package 
\cite{cmbfast} which is, in turn, based on the COSMICS package 
\cite{cosmics}. 
The evolution of the scale factor is integrated numerically from the usual Friedmann-Lema\^itre equations
\begin{equation}
{\mathcal H}^2 = \frac{8\pi G}{3} a^2 \rho_{\mathrm{t}},\qquad 
 {\mathcal H}^2 - {\mathcal H}' = 4 \pi G a^2 ( p_{\mathrm{t}} + \rho_{\mathrm{t}}),
\end{equation}
where $\rho_{\mathrm{t}}$ and $p_{\mathrm{t}}$ are, respectively, the total 
energy density and pressure of the plasma.
Equation (\ref{E3}) will then be supplemented by the  governing 
equations for the magnetic fields as well as by the  CDM particles and by the neutrino component:
\begin{eqnarray}
&&\delta_{\mathrm{c}}' = - \theta_{\mathrm{c}} + \frac{h'}{2},\hspace{2cm}
\theta_{\mathrm{c}}' + {\mathcal H} \theta_{\mathrm{c}} =0.
\label{CDMS1}\\
&&\delta_{\nu}' = -\frac{4}{3} \theta_{\nu} + \frac{2}{3} h',
\hspace{1.5cm}
\theta_{\nu}' = \nabla^2 \sigma_{\nu}  - \frac{1}{4} \nabla^2\delta_{\nu},
\label{nu1}\\
&& \sigma_{\nu}' = \frac{4}{15} \theta_{\nu}  - \frac{2}{15} h' - \frac{4}{5} \xi',
\label{nu2}
\end{eqnarray}
where $\sigma_{\nu} = {\mathcal F}_{\nu 2}/2$ is 
the quadrupole of the perturbed phase space distribution. Higher multipoles of the Boltzmann hierarchy (like 
the octupole ${\mathcal F}_{\nu  3}$)
 follow by setting initial conditions on the lower multipoles. The governing equations for baryons and photons are given by Eq. (\ref{E3}) together with
\begin{equation}
\theta_{\gamma}' = -\frac{1}{4} \nabla^2 \delta_{\gamma} + \epsilon' (\theta_{\mathrm{b}} - \theta_{\gamma}),\hspace{0.8cm}  \delta_{\gamma}' = - \frac{4}{3} \theta_{\gamma} + \frac{2}{3} h',
\hspace{0.8cm} \delta_{\mathrm{b}}' = - \theta_{\mathrm{b}} + \frac{h'}{2} + \frac{\vec{E}\cdot (\vec{\nabla} \times \vec{B})}{4\pi \sigma a^4 \rho_{\mathrm{b}}},
\label{PHOT}
\end{equation}
where the last term in the evolution of $\delta_{\mathrm{b}}$ is negligible at finite conductivity. 
The plasma and the magnetic fields all 
gravitate and contribute to the Hamiltonian and momentum constraints whose specific form is, respectively:
\begin{eqnarray}
&& 2\nabla^2\xi + {\mathcal H} h' = - 8\pi Ga^2 [\delta_{\mathrm{s}}\rho_{\mathrm{t}} + \delta\rho_{\mathrm{B}}],\qquad 
\label{HAM1}\\
&& \nabla^2 \xi' =  4\pi G a^2\biggl\{ ( p_{\mathrm{t}} + \rho_{\mathrm{t}}) \theta_{\mathrm{t}} + \frac{\vec{\nabla}\cdot [\vec{J} \times \vec{B}]}{4\pi a^4 \sigma}\biggr\},
 \label{MOM1}
 \end{eqnarray}
where $\delta_{\mathrm{s}}\rho_{\mathrm{t}}$ is the total density 
fluctuation in the $S$ gauge and $(p_{\rm t} + \rho_{\rm t}) \theta_{\rm t} = \sum_{{\rm a}} (p_{\rm a} + \rho_{\rm a})
\theta_{\rm a}$ is the total peculiar velocity.
Since the conductivity $\sigma$ is always large, the contribution of the MHD Poynting  vector is, in practice, always 
negligible.  The $(ij)$ components 
of the perturbed Einstein equations read
 \begin{eqnarray}
 && h'' + 2 {\mathcal H} h' + 2 \nabla^2 \xi = 24 \pi Ga^2 [\delta p_{\mathrm{t}} + \delta p_{\mathrm{B}}],
\label{SP1}\\
&& (h + 6 \xi)'' + 2 {\mathcal H} ( h + 6 \xi)' + 2 \nabla^2 \xi = 24 \pi G a^2 [ (p_{\nu} + \rho_{\nu}) \sigma_{\nu} + 
(p_{\gamma} + \rho_{\gamma}) \sigma_{\mathrm{B}}],
\label{SP2}
\end{eqnarray}
where $\delta_{\mathrm{s}} p_{\mathrm{t}}$ is the fluctuation 
of the total pressure. In MHD, $4\pi \vec{J} = \vec{\nabla} \times \vec{B}$ so that the Lorentz force and the magnetic anisotropic stress (associated with 
$\sigma_{\mathrm{B}}$) are related by:
 \begin{equation}
 \nabla^2 \sigma_{\rm B} = 
 \frac{3}{16\pi a^4 \rho_{\gamma}} \vec{\nabla}\cdot [
  (\vec{\nabla}\times \vec{B})
 \times \vec{B}] + 
 \frac{\nabla^2 \Omega_{\rm B}}{4},\qquad \Omega_{\rm B}(\vec{x}) = \frac{\delta\rho_{\rm B}(\tau, \vec{x})}{\rho_{\gamma}(\tau)}.
 \label{magndef}
 \end{equation}
The simplest set  of initial conditions to be imposed on the hierarchies of the fluctuations in the 
intensity and of the polarization is the magnetized adiabatic mode. 
 When  Coulomb and Thompson couplings are both 
tight (i.e. $\theta_{\mathrm{b}} \simeq \theta_{\gamma} = \theta_{\gamma\mathrm{b}}$)
the whole system of the governing equations can be solved in the limit 
when the relevant wavelengths are larger than the Hubble radius prior to matter-radiation equality (i.e., in terms of the wave-number $k$, $k\tau \ll 1$).  To lowest order in $k\tau$ the magnetized adiabatic mode reads, in Fourier space:
\begin{eqnarray}
 \xi(k,\tau) &=& - 2 C(k) + \biggl[\frac{4 R_{\nu} + 5}{6 ( 4 R_{\nu} + 15)} C(k) + \frac{R_{\gamma} ( 4 \sigma_{\mathrm{B}}(k) - R_{\nu} \Omega_{\mathrm{B}}(k))}{ 6 ( 4 R_{\nu} + 15)}\biggr] k^2 \tau^2,
\label{S1}\\
h(k,\tau) &=& - C(k) k^2 \tau^2 - \frac{1}{36} \biggl[ \frac{8 R_{\nu}^2 - 14 R_{\nu} - 75}{(2 R_{\nu} + 25)(4 R_{\nu} + 15)} C(k) 
\nonumber\\
&+& \frac{R_{\gamma} ( 15 - 20 R_{\nu})}{10( 4 R_{\nu} + 15) ( 2 R_{\nu} + 25)} (R_{\nu}\Omega_{\mathrm{B}}(k) - 4 \sigma_{\mathrm{B}}(k))\biggr] k^4 \tau^4,
\label{S2}\\
\delta_{\gamma}(k,\tau) &=& - R_{\gamma} \Omega_{\mathrm{B}}(k) - \frac{2}{3} \biggl[ C(k) - \sigma_{\mathrm{B}}(k) + \frac{R_{\nu}}{4} \Omega_{\mathrm{B}}(k)\biggr] k^2 \tau^2,
\label{S3}\\
\delta_{\nu}(k,\tau) &=& - R_{\gamma} \Omega_{\mathrm{B}}(k) - \frac{2}{3} \biggl[ C(k) + \frac{R_{\gamma}}{4 R_{\nu}}\biggl( 4\sigma_{\mathrm{B}}(k) - R_{\nu} \Omega_{\mathrm{B}}(k)\biggr)\biggr] k^2 \tau^2,
\label{S4}\\
\delta_{\mathrm{c}}(k,\tau) &=& - \frac{3}{4}R_{\gamma} \Omega_{\mathrm{B}}(k) - \frac{C(k)}{2} k^2 \tau^2,
\label{S5}\\
\delta_{\mathrm{b}}(k,\tau) &=& - \frac{3}{4}R_{\gamma} \Omega_{\mathrm{B}}(k) - \frac{1}{2} \biggl[ C(k) - \sigma_{\mathrm{B}}(k)+ \frac{R_{\nu}}{4} \Omega_{\mathrm{B}}(k) \biggr] k^2\tau^2,
\label{S6}\\
\theta_{\gamma\mathrm{b}}(k,\tau) &=& \biggl[ \frac{R_{\nu}}{4} \Omega_{\mathrm{B}}(k) - \sigma_{\mathrm{B}}\biggr] 
k^2 \tau -\frac{1}{36} \biggl[ 2 C(k) + \frac{R_{\nu} \Omega_{\mathrm{B}}(k) - 4 \sigma_{\mathrm{B}}(k)}{2}\biggr] k^4\tau^3,
\label{S7}\\
\theta_{\nu}(k,\tau) &=& \biggl[ \frac{R_{\gamma}}{R_{\nu}} \sigma_{\mathrm{B}}(k) - \frac{R_{\gamma}}{4} \Omega_{\mathrm{B}}(k)\biggr] k^2 \tau
- \frac{1}{36}\biggl[\frac{2 ( 4 R_{\nu} + 23)}{4 R_{\nu} + 15} C(k) 
\nonumber\\
&+& \frac{R_{\gamma}( 4 R_{\nu} + 27)}{2 R_{\nu} ( 4 R_{\nu} + 15)}( 4 \sigma_{\mathrm{B}}(k) - R_{\nu} \Omega_{\mathrm{B}}(k))\biggr] k^4 \tau^3,
\label{S8}\\
 \theta_{\mathrm{c}}(k,\tau) &=& 0,
\label{S9}\\
\sigma_{\nu}(k,\tau) &=& - \frac{R_{\gamma}}{R_{\nu}} \sigma_{\mathrm{B}}(k) + \biggl[ \frac{4 C(k)}{3( 4 R_{\nu} + 15)} + \frac{R_{\gamma}( 4 \sigma_{\mathrm{B}}(k) - R_{\nu} \Omega_{\mathrm{B}})}{ 2 R_{\nu}(4 R_{\nu} + 15)}\biggr] k^2 \tau^2,
\label{S10}
\end{eqnarray}
where $R_{\nu} = r/(r +1)$ with $r= 0.681(N_{\nu}/3)$ (and $R_{\gamma} = 1 - R_{\nu}$).
\begin{figure}
\begin{center}
\begin{tabular}{|c|c|}
      \hline
      \hbox{\epsfxsize = 7 cm  \epsffile{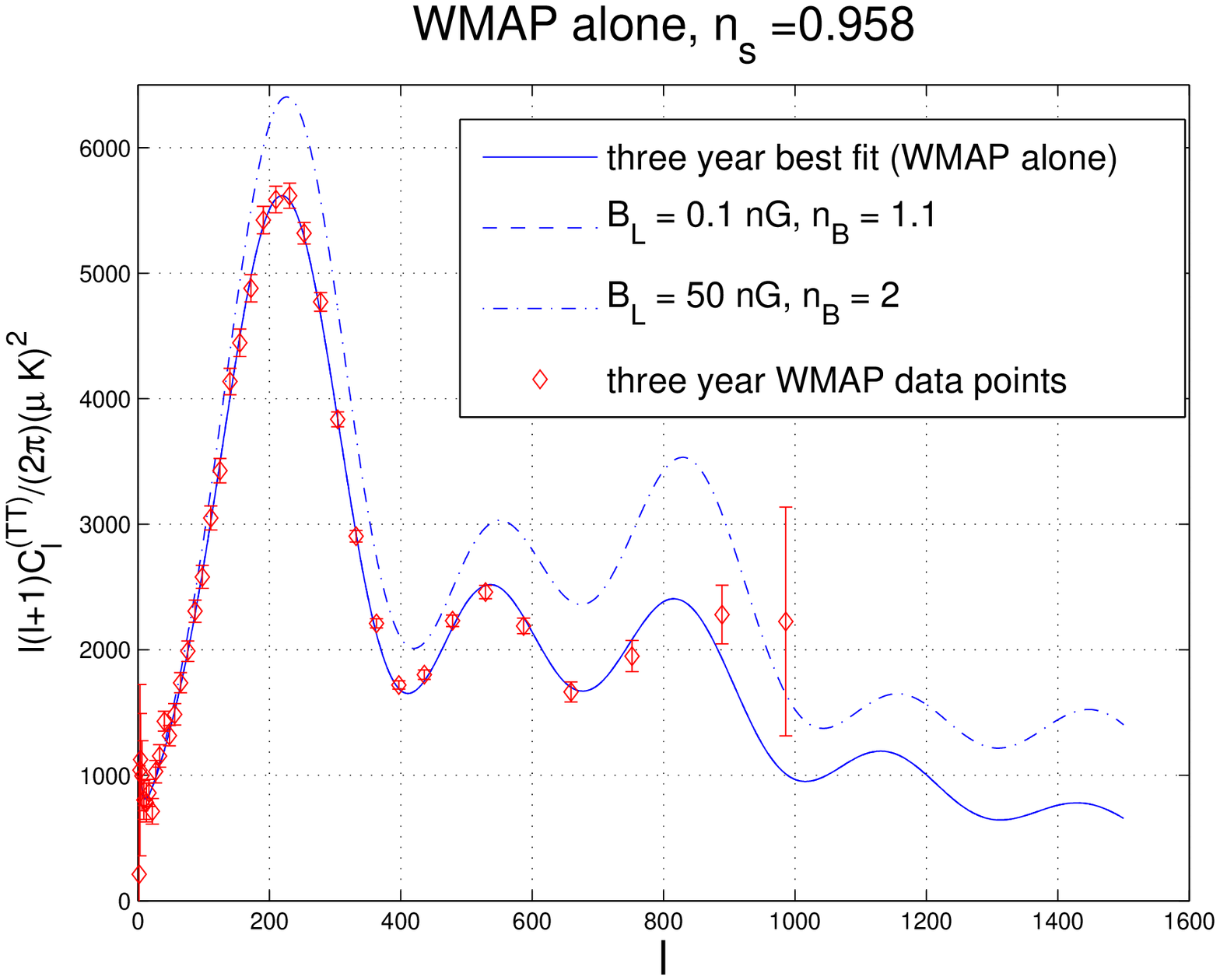}} &
     \hbox{\epsfxsize = 7 cm  \epsffile{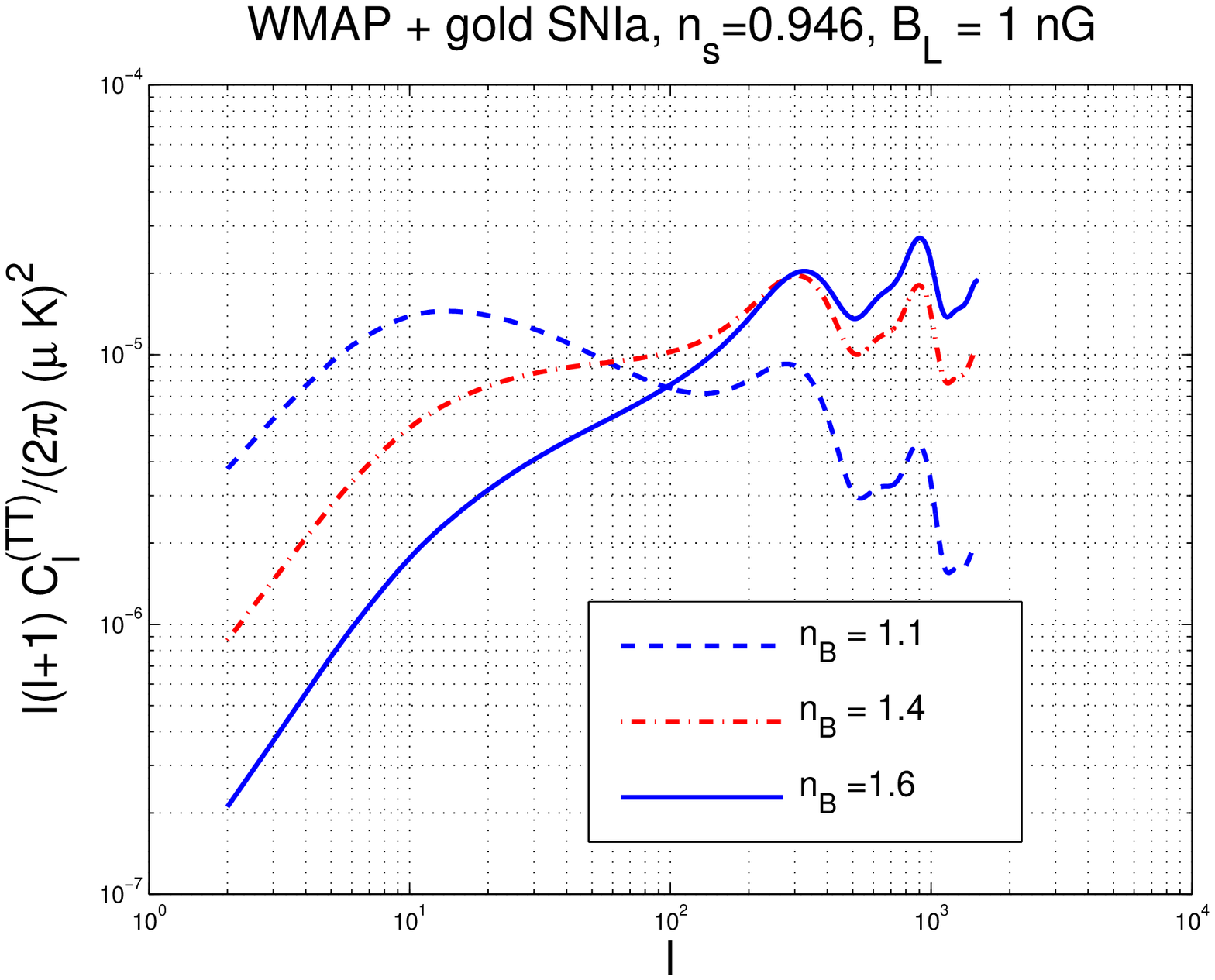}}\\
      \hline
\end{tabular}
\end{center}
\caption[a]{The magnetized temperature autocorrelations 
stemming from the magnetized adiabatic mode (left plot) 
and in the absence of an adiabatic component (plot at the right).
The other parameters are fixed to the central values of the best fit of WMAP three year data alone (plot at the left) and to the central values of the 
best fit of the WMAP data supplemented by the gold sample of supernovae (see \cite{data}).}
\label{Figure1}
\end{figure}
The constant mode of $h$ leads to a gauge mode and should be projected out of the solution. The second 
potentially dangerous gauge mode is fixed by setting to $0$ the CDM peculiar velocity $\theta_{\mathrm{c}}$.
The spectrum of $C(k)$ is simply related to the spectrum of ${\mathcal R}$ which parametrizes 
the curvature perturbations on comoving orthogonal hypersurfaces \cite{mg}. Deep in the radiation epoch, according to Eq. (\ref{S1}), ${\mathcal R}(k,\tau) = - 2 C(k)$. Consequently 
the initial conditions for the adiabatic component will be given in terms of the spectrum of ${\mathcal R}$, i.e.  ${\mathcal P}_{{\mathcal R}}(k) ={\mathcal A}_{\mathcal R} (k/k_{\mathrm{p}})^{n_{\mathrm{s}}-1}$ where $n_{\mathrm{s}}$ is the adiabatic spectral index and ${\mathcal A}_{{\mathcal R}}$ is 
the amplitude of the scalar power spectrum at the pivot scale $k_{\mathrm{p}}=0.002\, \mathrm{Mpc}^{-1}$. The spectrum of the magnetized contribution will be encoded in the spectra of 
$\sigma_{\mathrm{B}}$ and 
$\Omega_{\mathrm{B}}$, i.e. respectively, 
${\cal P}_{\sigma}(k) 
 = {\mathcal G}(n_{\mathrm{B}})\overline{\Omega}_{{\rm B}\, L}^2 
 (k/k_{\mathrm{L}})^{2 (n_{\sigma}-1)}$ and 
${\cal P}_{\Omega}(k) = {\mathcal F}(n_{\mathrm{B}}) \overline{\Omega}_{{\rm B}\, L}^2 (k/k_{\mathrm{L}})^{2 (n_{\mathrm{B}}-1)}$
where $k_{\mathrm{L}}$ denotes the magnetic pivot scale\footnote{The range of magnetic spectral indices discussed here is $1< n_{\mathrm{B}} < 5/2$. In this case, regularizing the magnetic field with a Gaussian 
window function, the spectra of the $\Omega_{\mathrm{B}}$ and $\sigma_{\mathrm{B}}$ are the ones 
reported in Eq. (\ref{CORR11}). For values outside the mentioned range cut-offs are required either in the infra-red or in the ultra-violet
\cite{kmg}. The value of the magnetic pivot scale implies that $B_{\mathrm{L}}$ coincides 
effectively with the proper amplitude of the magnetic field regularized over a Mpc window at the onset 
of protogalactic collapse \cite{mg}.} which will be taken $1\,{\mathrm{Mpc}}^{-1}$. In the minimal 
m$\Lambda$CDM scenario $n_{\sigma} = n_{\mathrm{B}}$.
The spectra ${\mathcal P}_{\Omega}(k)$ and ${\mathcal P}_{\sigma}(k)$ 
are not arbitrary but rather computed from Eq. (\ref{E1}) through a rather standard procedure (see, for details, \cite{kmg}):
\begin{eqnarray}
&&\overline{\Omega}_{\mathrm{BL}} = \frac{B_{\mathrm{L}}^2}{8\pi \overline{\rho}_{\gamma}} = 7.5 \times 10^{-9} \biggl(\frac{B_{\mathrm{L}}}{\mathrm{nG}}\biggr)^{2},
\nonumber\\
&&{\mathcal G}(n_{\mathrm{B}}) = \frac{(2\pi)^{2(n_{\mathrm{B}} -1)}}{\Gamma^2\biggl(\frac{n_{\mathrm{B}}-1}{2}\biggr)}\biggl[ \frac{ n_{\mathrm{B}} + 29}{15 ( 5 - 2 
n_{\mathrm{B}})( n_{\mathrm{B}} -1)}\biggr],\hspace{1cm}{\mathcal F}(n_{\mathrm{B}}) = 
\frac{20 ( 7 - n_{\mathrm{B}})}{n_{\mathrm{B}} + 29} {\mathcal G}(n_{\mathrm{B}}).
\label{CORR11}
\end{eqnarray}
\begin{figure}
\begin{center}
\begin{tabular}{|c|c|}
      \hline
      \hbox{\epsfxsize = 7 cm  \epsffile{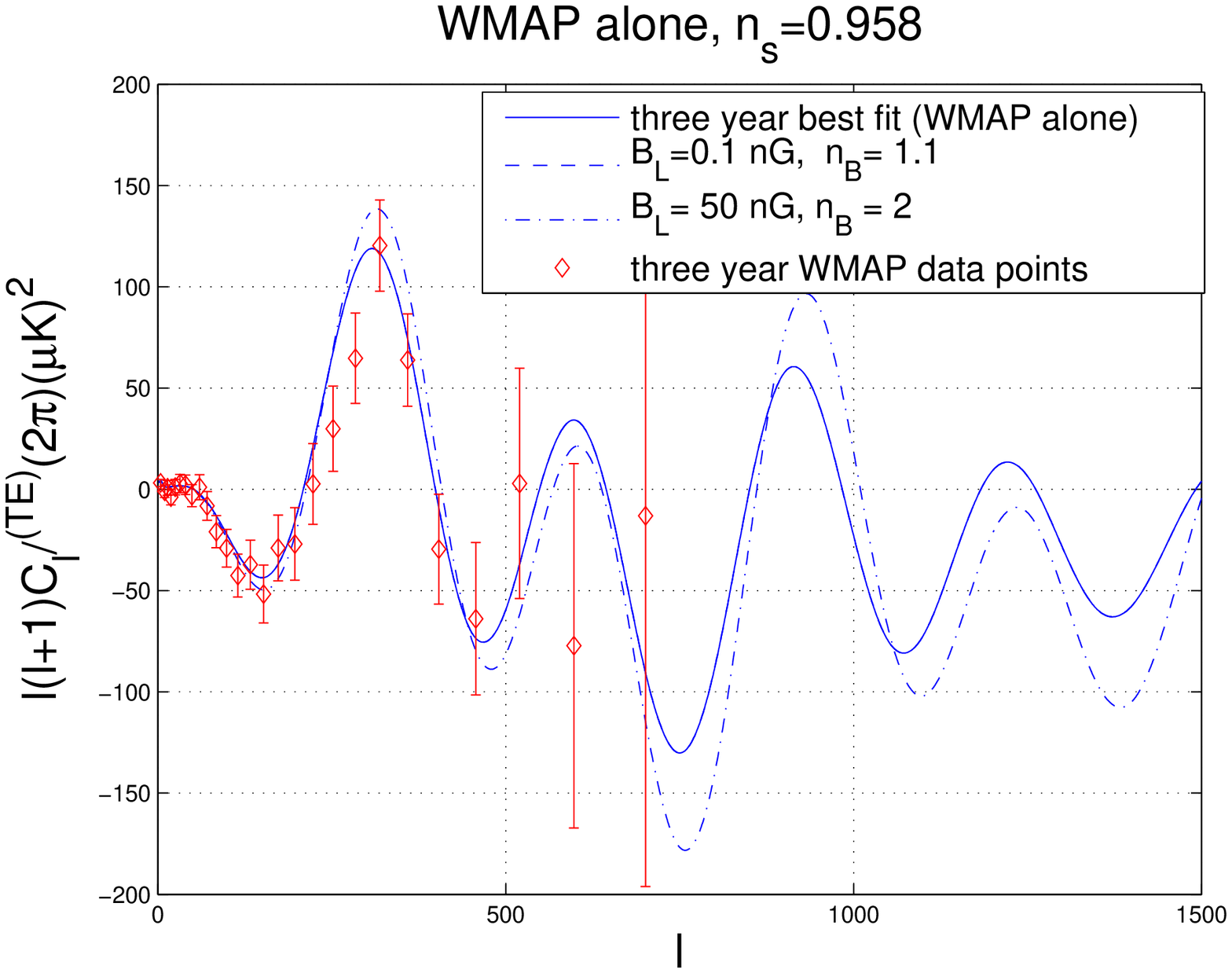}} &
     \hbox{\epsfxsize = 7 cm  \epsffile{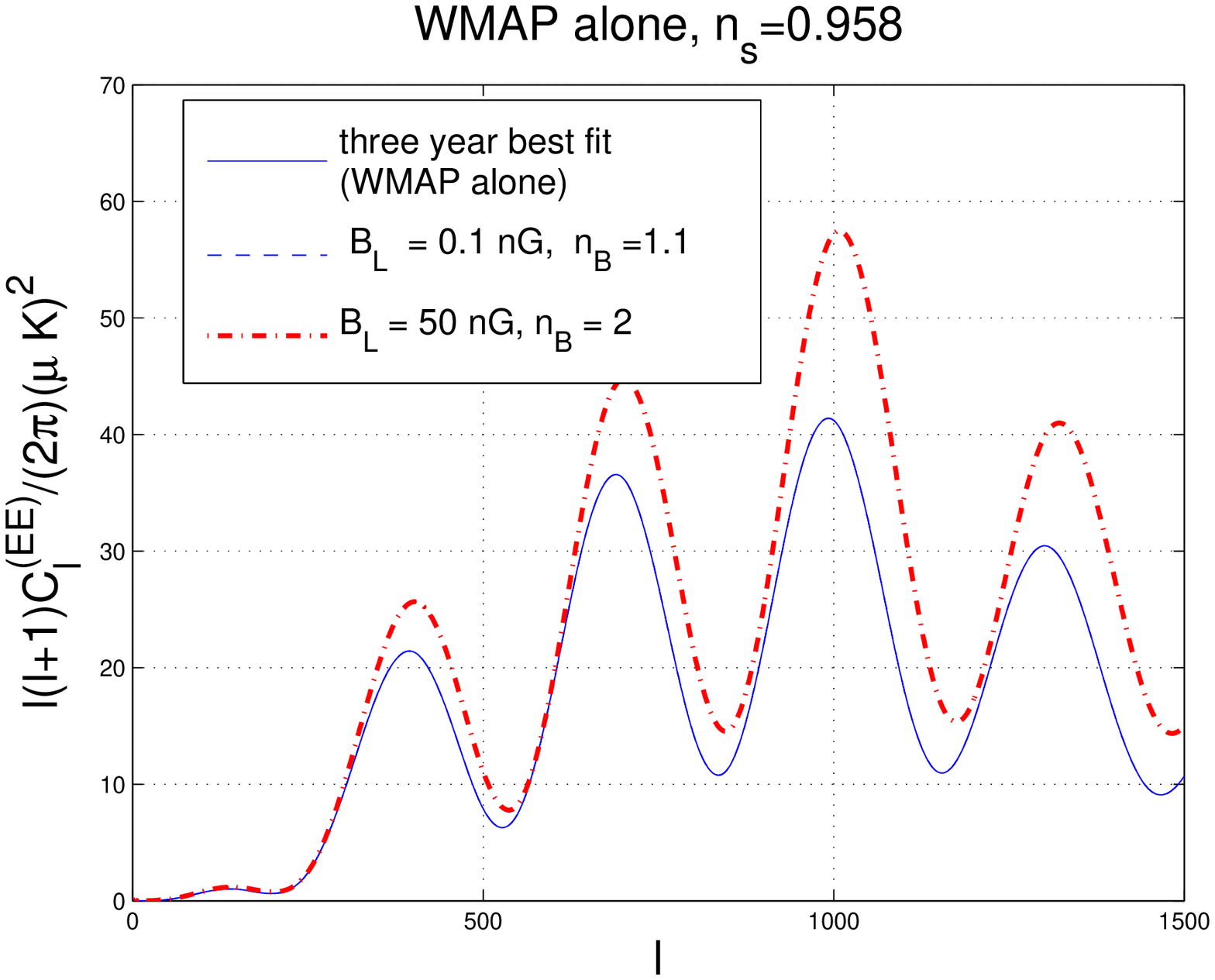}}\\
      \hline
\end{tabular}
\end{center}
\caption[a]{The magnetized TE and EE correlations are illustrated for different values of the spectral indices and different values of the magnetic field intensities.}
\label{Figure2}
\end{figure}
In Fig. \ref{Figure1}(plot at the left) the effects of the magnetized contribution on the temperature autocorrelations is illustrated for different values of the spectral index and of the magnetic field intensity.  
If the strength of the magnetic field 
is augmented (or the value of the spectral index becomes bluer) 
a distortion of the second  and of the third peaks is correlated with an increase of the first peak. 
This spectral distortion fits with the results of a  semi-analytical 
calculation conducted in the longitudinal gauge (see \cite{mg}, last 
reference). In spite of the correct spectral distortion on the shape,
the semi-analytical argument is intrinsically less accurate. 
The predicted height of the first peak varies in  a non-monotonic 
way with the variation of the spectral index and of the magnetic field 
intensity \cite{kmg}.  If the adiabatic mode is absent from the initial conditions (Fig. \ref{Figure1} plot at the right) the amplitude of the TT correlations is always much smaller than in the case of the magnetized adiabatic mode. Furthermore a hump appears at intermediate multipoles.
In Fig. \ref{Figure2} we illustrate the angular power spectrum of the 
temperature-polarization cross-correlations (for short the TE correlations) as well as the 
angular power spectrum for the polarization autocorrelations (for short the EE correlations). According to Fig. \ref{Figure3}, for high $\ell$ (where, hopefully, there will be, in the near future more precise data from the Planck explorer mission) the TT correlation 
is definitely sensitive, for $\ell \gg 1500$, to a nG magnetic field. The TE 
correlation is also sensitive to the nG range (see Fig. \ref{Figure3}, plot at the right) especially 
as soon as the spectral index increases from the nearly scale-invariant limit.
\begin{figure}
\begin{center}
\begin{tabular}{|c|c|}
      \hline
      \hbox{\epsfxsize = 7.5 cm  \epsffile{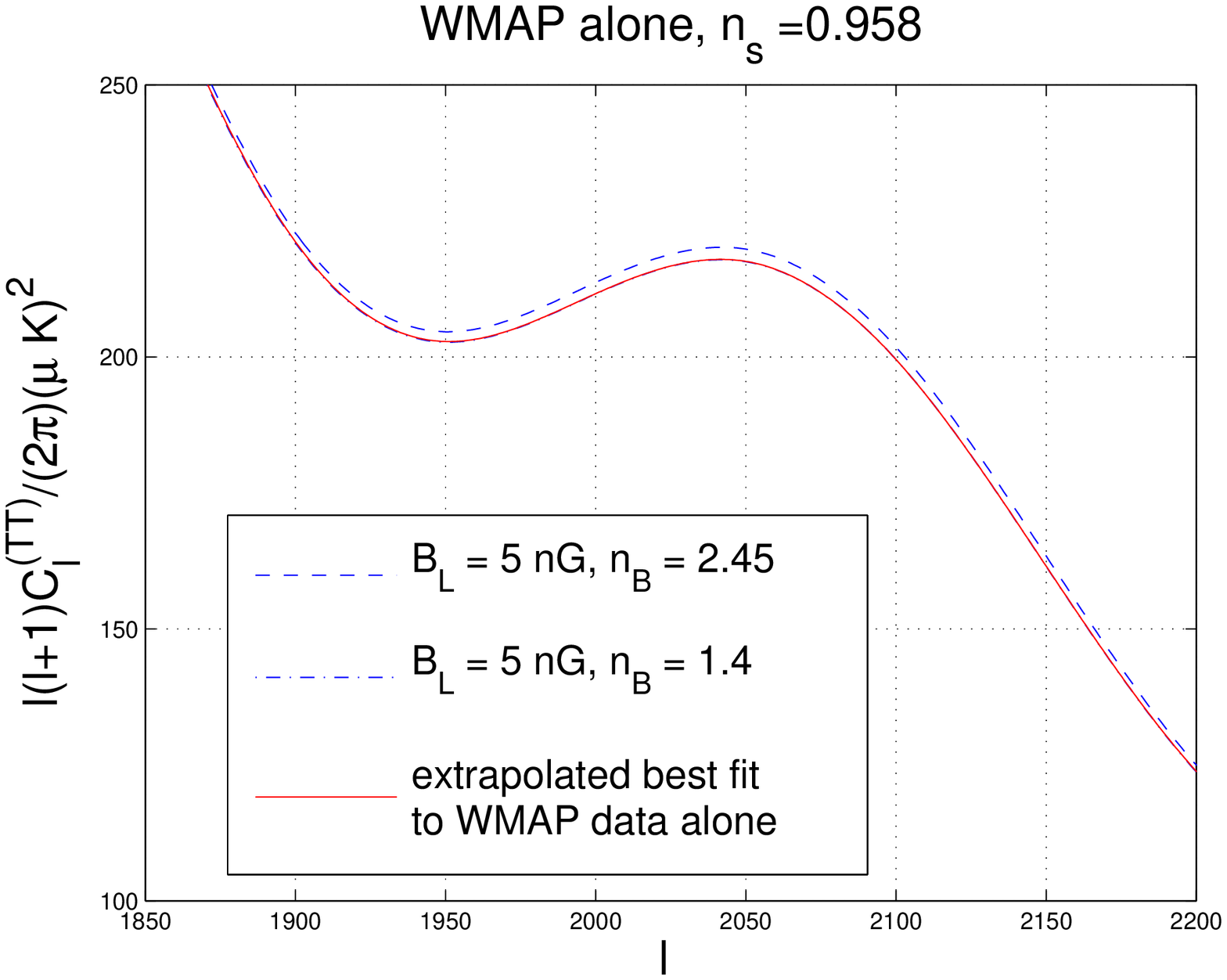}} &
     \hbox{\epsfxsize = 7.5 cm  \epsffile{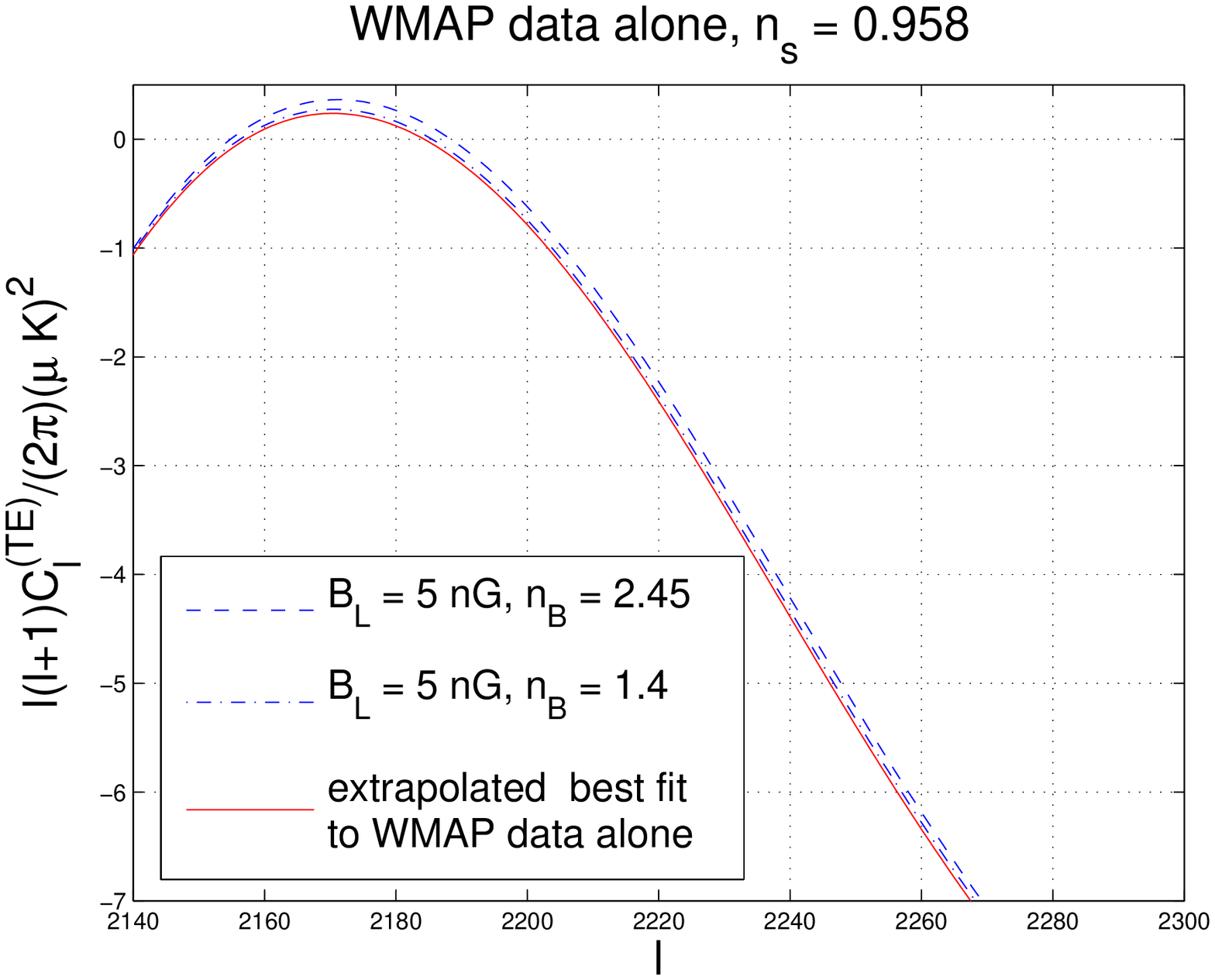}}\\
      \hline
\end{tabular}
\end{center}
\caption[a]{The TT and TE correlation for $\ell \gg 1500$. The full curve 
is the extrapolation of the three year best fit of the WMAP data alone.}
\label{Figure3}
\end{figure}
The magnetized CMB 
anisotropies have never been computed consistently and systematically for the scalar 
modes of the geometry which are the ones observationally more relevant.  In this paper  we built a consistent numerical approach to the problem and presented a magnetized completion of the
$\Lambda$CDM paradigm.
Our approach is accurate enough to resolve nG magnetic fields.
The usual custom of setting a bound on the magnetic field becomes too simplistic in the 
light of the present results and according to the standards of CMB physics. In a  similar perspective, for instance, the 
bound on the ratio between tensor and scalar power spectra can only 
stem from an appropriate strategy of parameter extraction where the tensor contribution 
is included in the CMB anisotropy calculations and fits. The same must be done in the case of the m$\Lambda$CDM 
scenario. If the correct strategy is not enforced, potentially interesting 
degeneracies between the parameters of the magnetized background and other cosmological parameters can be totally overlooked.
Consequently, our program is, in the short 
run, to extend and complete our code to the case of tensor and vector modes (which are known to be far less relevant at large scales) and to the case of the various magnetized isocurvature modes.
In parallel we ought to test the m$\Lambda$CDM scenario against 
the various data sets eagerly waiting for Planck data and its claimed high accuracy for 
large multipoles.

K.E.K. is supported by the ``Ram\'on y Cajal''  program and grants FPA2005-04823 and FIS2006-05319 of the Spanish Science Ministry.

\end{document}